\documentclass[10pt,journal,romanappendices]{IEEEtran}

\usepackage[tight,footnotesize]{subfigure}

\usepackage{graphicx}
\usepackage{amssymb}
\usepackage{array}
\usepackage{dsfont} 

\usepackage{cite}
\usepackage{amsmath}
\usepackage{amsthm}
\usepackage{graphicx}
\usepackage{amssymb}

\usepackage{array}
\usepackage{color}
\usepackage{multirow}
\usepackage{ifpdf}
\usepackage[T1]{fontenc}
\graphicspath{{./Figs/}}                   

\theoremstyle{definition}

\newtheorem{theorem}{Theorem}

\newcommand{\prob}{\mathbb{P}}

\newcommand{\Ex}{\mathbb{E}}
\newcommand{\ve}[1]{\boldsymbol{#1}}
\newcommand{\R}{\mathbb{R}}

\newcommand{\defeq}{\triangleq}

\newcommand{\Hip}{\mathcal{H}}
\newcommand{\N}{\mathcal{N}}

\begin{document}
\title{Cooperative Spectrum Sensing Schemes with Partial Statistics Knowledge}

\author{{Juan Augusto Maya,~\IEEEmembership{Student Member,~IEEE,} Leonardo Rey Vega,~\IEEEmembership{Member,~IEEE,} and Cecilia G. Galarza}
\thanks{This work has been submitted to the IEEE for possible publication. Copyright may be 
transferred without notice, after which this version may no longer be accessible.}
\thanks{The three authors are with the University of Buenos Aires, Buenos Aires, Argentina. L. Rey Vega and C. Galarza are also with the CSC-CONICET, Buenos Aires, Argentina. 
Emails:  \{jmaya, lrey, cgalar\}@fi.uba.ar. This work was partially supported by the Peruilh grant and the projects UBACYT 20020130100751BA and CONICET PIP 112 20110100997.} %
}
\maketitle

\begin{abstract}
In this letter, we analyze the problem of detecting spectrum holes in cognitive radio systems. We consider that a group of unlicensed users can sense the radio signal energy, perform some simple processing and transmit the result to a central entity, where the decision about the presence or not of licensed users is made. We show that the proposed cooperative schemes present good performances even without any knowledge about the measurements statistics in the unlicensed users and with only partial knowledge of them in the central entity.
\end{abstract}


\section{Introduction}
Wireless spectrum is a vital resource required for radio communications. Worldwide, radio frequency bands are statically assigned to the licensed holders of the bands in large geographic areas. Within this paradigm, the wireless spectrum is commonly a scarce resource in high demand for current and future technologies. However, measurements campaigns have suggested that much of the licensed spectrum is frequently under-utilized in vast areas at different times.
In the recent years, cognitive radio (CR) systems has emerged as a possible solution for the spectrum shortage (see \cite{2015PoorSpectrumSensing} and references therein). In CR systems, unlicensed, or secondary users (SU), sense the spectrum in a particular place and time and wish to detect the presence or absence of the licensed, or primary users (PU), in order to use the spectrum when it is available. Although each SU may sense the spectrum alone and makes a decision, the typical low signal-to-noise ratio (SNR) of the PU signal at the SU receiver makes difficult to develop reliable detection schemes. Additionally, the so-called hidden-terminal problem arises in environments with shadowing fading: a SU could receive an undetectable weakly shadowed signal from the PU, decide to transmit and produce interference to the primary receiver. A way to tackle these issues is to allow SUs cooperating with each other to detect the PU signal. While some of the SUs receivers could be shadowed from the PU signal, it is unlikely that all of them are in a deep shadow simultaneously.

\begin{figure}[h]
\centering
\includegraphics[width=\linewidth]{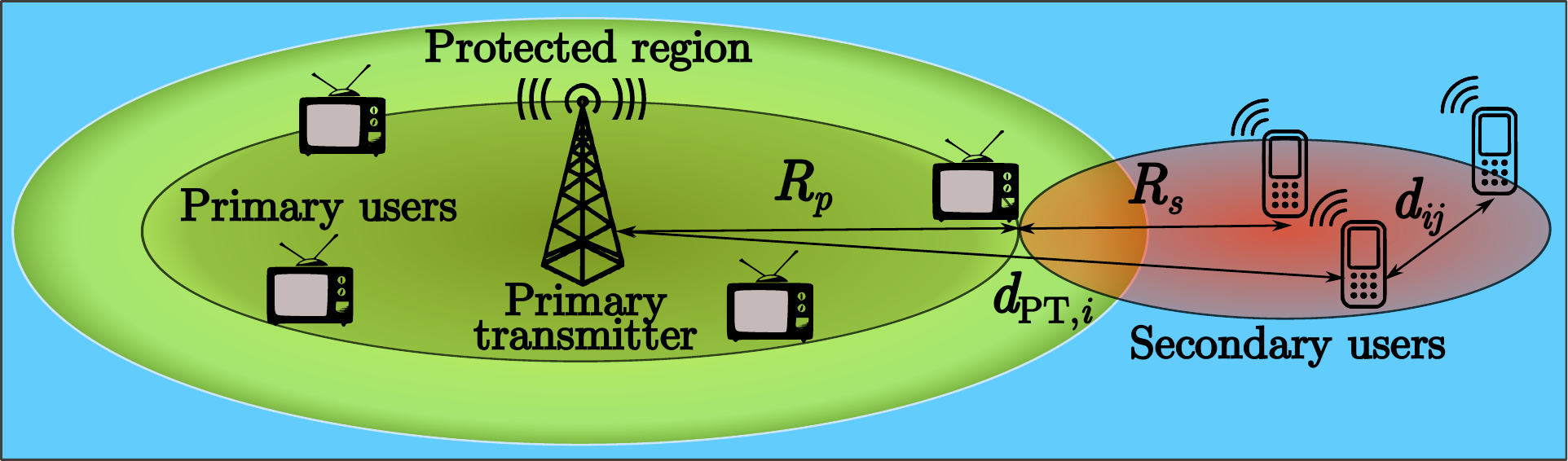}
\caption{A CR network with spectrum sensing devices. The SUs should be able to reliable detect far away activity (low SNR signals) of primary users.}
\label{fig:cr}
\end{figure}

The distributed detection problem with cooperative sensors for CR systems has been considered previously in the literature \cite{2015PoorSpectrumSensing,Veeravali2008CoopSensing}. However, most of the works assume total or partial knowledge of the measurements joint statistics, something difficult to achieve in a practical scenario.
Considering that the SUs in general do not know the transmission scheme used by the PUs, and the lack of training sequences for synchronization, demodulation of the PU signal is unfeasible. Therefore, non-coherent energy detectors are typically used in the SUs receivers. 
In Fig. \ref{fig:cr}, we show a possible scenario of a CR. A primary transmitter communicates with its primary receivers located inside the primary range $R_p$, defined as the maximum distance between a primary transmitter and a primary receiver.
The SUs sense the spectrum and decide if they are out of the \emph{protected region} of the primary system such that they can use the spectrum without causing harmful interference to the PUs. 

In this letter, we propose two simple distributed detection schemes based on the generalized log-likelihood ratio (GLLR) test where no prior knowledge of the measurements statistics is necessary. In both schemes, the SUs gather a set of measurements delivered by their non-coherent energy detectors, perform some processing and transmit the analog\footnote{A quantization scheme will be required in practice but its analysis is out of the scope of this letter.} result through a noiseless communication channel to the fusion center (FC). The FC makes a decision about the state of the PU through a threshold test, and then, broadcasts it to the SUs. The FC could be a dedicated processing terminal or even an ordinary SU.  
In the first scheme, each SU computes the \emph{quadratic mean} of the set of measurements gathered and transmits the result to the FC. The FC computes the average of all the signal transmitted by the SUs and makes a decision. This scheme is suitable for urban cells where the signal propagates through environments with dense scattering objects producing a severe shadowing fading.
In the second scheme, each SU computes the \emph{linear mean} of the set of energy measurements of the radio signal and transmits it to the FC. The FC averages out all the square of the received signals from the SUs and makes a decision. This scheme is suitable for rural environments where the amount of scattering objects in the signal path is low and the shadowing effect is weak.
Next, we present the signal model, derive the statistics and compute their performances. Finally, we analyze the results and elaborate the conclusions of this letter.

\section{Signal Model}
The basic task of the FC is to decide if the SUs are located inside the protected region or not (see Fig. \ref{fig:cr}) based on the measurements delivered by their energy detectors. Therefore, we have a binary hypothesis testing problem. Under $\Hip_1$, we consider that the primary transmitter is ON and the SUs are inside the protected region. In this case, the energy that the SUs measure will correspond to the superposition of the primary signal, some possible background interference from other networks and the SU receivers self noise. We assume that the primary signal is affected by both path loss and shadowing fading effects. The power received in a given SU $P_r$ (in dBm) separated from the primary transmitter by a distance $d$ is modeled by \cite{goldsmith2005wireless}
\begin{equation}
\begin{array}{l}
P_r(d) = P_t + K_\text{dB} - 10\gamma\log_{10}(d/d_0) - \psi_\text{dB}
\end{array}\end{equation}   
where $P_t$ (in dBm) is the transmitted power, $K_\text{dB}$ is a unit less constant that depends on the antenna characteristics and the average channel attenuation, $\gamma$ is the path loss exponent, $d_0$ is a reference distance for the antenna far field, and $\psi_\text{dB}$ is a random variable that models the shadowing fading effect. We adopt the popular log-normal model for the shadowing fading meaning that the random variable expressed in dB $\psi_\text{dB}$ is a zero-mean Gaussian random variable with variance $\sigma^2_\text{SH}$, which depends on the propagation environment and is assumed to be known in the FC. In a wireless scenario, the signal power measured by two receivers close each other presents certain correlation. We model such correlation to decay exponentially with the distance between them as in \cite{goldsmith2005wireless}. Thus, the autocorrelation function is $R(d)= \sigma^2_\text{SH} e^{-d/d_c}$ where $d$ is the distance between two SUs and $d_c$ is the decorrelation distance.

Under $\Hip_0$, when the primary transmitter is OFF or the SUs are outside the protected region, we assume that the energy received by the SUs is dominated by the noise in their electric circuits. Therefore, we assume that the measurements across the SUs are independent. Moreover, in practical scenarios, it is difficult to estimated the noise power due to calibration errors as well as changes in thermal noise caused by temperature variations. Thus, the noise power can only be known with a certain degree of accuracy and we model this situation assuming that the measurements delivered by the energy detectors are log-normally distributed with mean $m_0$ and variance $\sigma_0^2$ \cite{Veeravali2008CoopSensing}. Both parameters depend on the energy detector characteristics and are assumed to be known. 

In order to improve the detection performance in the FC, each SU can take several measurements across the time. It is clear that the reliability of the detection scheme can be increased with the number of measurements although this will be limited by the detection delay required for the application. 
In this work, we will assume that the measurements at each SU are independent and identically distributed (iid). Under $\Hip_1$, we can justify this assumption based on the fact that the small-scale fading, typically present in wireless communications, will produce uncorrelated fluctuations in the signal if the coherence time is of the order of the sampling time. Under $\Hip_0$, the measurements are dominated by the thermal noise which can be considered uncorrelated also in time. 

Consider that each SU can subtract the mean value $m_0$ (in dBm) to its set of energy measurements to obtain the set of observations $\{y_{ij}\}_{j=1}^m$, where the indexes $i$ and $j$ identify the SU, and the time slot, respectively, and $m$ is the number of measurements taken by each SU. 
Let $\ve y_j = [y_{j1},\dots,y_{jn}]^T$ and $\ve y =[\ve y_1^T,\dots,\ve y_m^T]^T$ be the vector of observations collected by $n$ SUs at the $j$-th time slot and the whole vector of measurements, respectively. The hypothesis testing problem consists in choosing $\Hip_0$ or $\Hip_1$:
\begin{equation}
\left\{
\begin{array}{ll}
\Hip_0: & \ve y \sim \N(\ve 0, \sigma_0^2 I_{nm}) \\
\Hip_1: & \ve y \sim \N(\ve 1 \otimes \ve \mu, I_{m}\otimes\Sigma_1),
\end{array}
\right.
\label{eq:test2}
\end{equation}
where $\N(\ve a,B)$ denotes the Gaussian distribution with mean $\ve a$ and covariance matrix $B$, $\otimes$ is the Kronecker product, $\ve 1$ is a column vector of $m$ ones, $\ve \mu =[\mu_1,\dots,\mu_n]^T$, with $\mu_i=P_r(d_{\text{PT},i})-m_0$ and $d_{\text{PT},i}$ is the distance between the primary transmitter (PT) and the $i$-th SU receiver. 
We assume that both the measurement inaccuracy and the shadowing effects are additive in the dB scale,  $\Sigma_1 = \sigma_0^2 I_{n} + \sigma^2_\text{SH}\Sigma_\text{SH} = \sigma^2_1 \tilde{\Sigma}_1$, where $I_n$ is the identity matrix of dimension $n$, $\Sigma_\text{SH}$ is the $n\times n$ normalized covariance matrix given by the shadowing fading, i.e., $(\Sigma_\text{SH})_{ij}=e^{-d_{ij}/d_c}$, and $d_{ij}$ is the distance between the $i$-th and $j$-th SU. We also let $\sigma^2_1=\sigma^2_0+\sigma^2_\text{SH}$ and $\tilde{\Sigma}_1$ be the variance and normalized covariance matrix of $\ve y_i$ under $\Hip_1$, respectively. 
Given that $\Sigma_\text{SH}$ depends on the distance between each pair of SUs, in practical situations it is difficult to know or estimate it. For circumventing this issue, we consider the GLLR test, which uses the maximum likelihood (ML) estimation of the unknown parameters to compute the statistic. 

\section{Statistics}
The (normalized) GLLR test compares with a predefined threshold $\tau$ the following statistic: $T_\textrm{GLLR}(\ve y) = \frac{1}{nm}\log\frac{p_1\left(\ve y;\hat{\ve \mu},\hat{\Sigma}_\text{1}\right)}{p_0\left(\ve y\right)}$,
where $p_j(\ve y )$ is the probability density function of $\ve y$ under $\Hip_j$, $j=0,1$ and $\hat{\ve \mu} = \tfrac{1}{m}\sum_{i=1}^{m} \ve y_i$ and $\hat{\Sigma}_\text{1}= \tfrac{1}{m}\sum_{i=1}^{m} (\ve y_i- \hat{\ve\mu})(\ve y_i- \hat{\ve\mu})^T$ are the ML estimators of the mean and the covariance matrix of $\ve y$ under $\Hip_1$, respectively. 
A CR system operating with licensed users must guaranty to work below a maximum probability of interference level $\beta$, while minimizing the probability of miss an opportunity to use the spectrum when it is free. 
Let $\prob_j(\cdot)$ be the probability measure under $\Hip_j$, $j=0,1$, let $\hat{\Hip}$ be the made decision and assume that the SUs transmit if they detect the spectrum to be free. The probability of interference is $P_\text{int}= \prob_1(\hat{\Hip}=\Hip_0)= \prob_1(T<\tau)$ and the miss opportunity probability is $P_\text{mo}=\prob_0(\hat{\Hip}=\Hip_1)= \prob_0(T>\tau)$, where the threshold $\tau$ is such that $P_\text{int}\leq \beta$.
Given that $\sigma^2_0$ and $\sigma^2_\textrm{SH}$ are assumed to be known, the GLLR statistic can be expressed as
\begin{equation}
\begin{array}{ll}
T_\textrm{GLLR}(\ve y)\hspace*{-3mm}&= \tfrac{1}{2}\log\tfrac{\sigma^2_0}{\sigma^2_1}-\tfrac{1}{2n}\log|\hat{\tilde{\Sigma}}_1|\\
&-\tfrac{1}{2nm\sigma^2_1}\sum_{i=1}^{m}
(\ve y_i\!-\!\hat{\ve\mu})^T \hat{\tilde{\Sigma}}_1^{-1} (\ve y_i\!-\!\hat{\ve\mu}) - \tfrac{\sigma^2_1}{\sigma^2_0} \|\ve y_i\|^2.
\label{eq:gllr2}
\end{array}
\end{equation}
To compute this statistic, each SU should employ $m$ channel uses\footnote{A channel use takes place when a symbol is transmitted.} to transmit to the FC its whole set of measurements, which is energy and bandwidth costly, two scarce resources if the SUs are battery powered mobile users. 
Instead, we propose to analyze two different scenarios and to derive different detection schemes for each one of them. The goal is to be able to transmit a summary of the set of measurements only.
Consider urban scenarios with rich scattering where $\sigma^2_\textrm{SH}\gg\sigma^2_0$ (which implies $\sigma^2_1\gg\sigma^2_0$). In this scenario, the second term of the second line of (\ref{eq:gllr2}) dominates the behavior of the statistic and it can be approximated by the quadratic mean (QM) statistic defined as 
\begin{equation}
T_\textrm{QM}(\ve y) = \tfrac{1}{nm}\sum_{i=1}^{m} \|\ve y_i\|^2= \tfrac{1}{n}\sum_{j=1}^{n} \left(\frac{1}{m}\sum_{i=1}^{m} y_{ij}^2\right).
\label{eq:QM}
\end{equation}
In this case, each SU computes its quadratic mean $\frac{1}{m}\sum_{i=1}^{m} y_{ij}^2$ and communicates it to the FC, where the final average is computed to obtain $T_\textrm{QM}(\ve y)$. 
On the other hand, in rural environments where the shadowing fading is weak, $\sigma^2_\textrm{SH}\ll\sigma^2_0$ (which implies $\sigma^2_1\approx\sigma^2_0$ and $\tilde{\Sigma}\approx I_n$), (\ref{eq:gllr2}) can be approximated by the linear mean (LM) statistic defined as
\begin{equation}
T_\textrm{LM}(\ve y) =  \tfrac{1}{n} \|\hat{\ve \mu}\|^2 = \frac{1}{n}\sum_{j=1}^{n} \left(\frac{1}{m}\sum_{i=1}^{m} y_{ij}\right)^2,
\label{eq:LM}
\end{equation}
where each SU computes the linear mean $\frac{1}{m}\sum_{i=1}^{m} y_{ij}$ and transmits it to the FC, which square and averages the received signals to get $T_\textrm{LM}(\ve y)$.
\section{Computation of the Error Probabilities}
The error probabilities can be computed using the following theorem \cite{Maya2015ESC}:
\vspace{-2mm}
\begin{theorem}[]
\label{thm:Pe}
Let $\{y_k\}_{k=1}^n$ be mutually independent random variables with logarithmic moment generating function (LMGF) $\mu_{k}(s)\defeq \log\Ex\left(e^{y_k s}\right)$. Assume that $\Ex(y_k^2)$ and $\Ex(|y_k-\Ex(y_k)|^3)$ exist and are finite  $\forall k\in[1:n]$. Let $T_n=y_1+\dots + y_n$ with LMGF $\mu_{T_n}\defeq \log\Ex\left(e^{T_n s}\right)$ and let $\tau_n\in \R$. Then, if $\tau_n> \Ex(T_n)$,  
$\prob(T_n>\tau_n) =
(\tfrac{1}{\sqrt{2\pi s_0^2\ddot{\mu}_{T_n}(s_0)}} + \mathcal{O}(\tfrac{1}{\sqrt{n}})) e^{-(s_0\dot{\mu}_{T_n}(s_0) - \mu_{T_n}(s_0))}$,
where $s_0>0$ satisfies $\tau_n = \dot{\mu}_{T_n}(s_0)$. Now, if $\tau_n < \Ex(T_n)$,  
$\prob(T_n < \tau_n) =(\tfrac{1}{\sqrt{2\pi s_1^2\ddot{\mu}_{T_n}(s_1)}} + \mathcal{O}(\tfrac{1}{\sqrt{n}})) e^{-(s_1\dot{\mu}_{T_n}(s_1) - \mu_{T_n}(s_1))}$, where $s_1<0$ satisfies $\tau_n = \dot{\mu}_{T_n}(s_1)$.
\end{theorem}

The statistics (\ref{eq:gllr2}), (\ref{eq:QM}) and (\ref{eq:LM}) can be expressed as $T(\ve y) = \ve y^T A \ve y + \ve y^T \ve b + c$, where the matrix $A$, the vector $\ve b$ and the scalar $c$ are properly chosen for each case. If $\ve y\sim \N(\ve m, C)$, the LMGF of $T(\ve y)$ is
$\mu_T(s)= -\tfrac{1}{2}\log|I-2sCA|-\tfrac{1}{2}\ve m^T C^{-1} \ve m + \tfrac{1}{2}(s\ve b+C^{-1}\ve m)^T (C^{-1} - 2sA)^{-1} (s\ve b+C^{-1}\ve m) + sc $ and its first and second derivatives $\dot{\mu}_T(s)$ and $\ddot{\mu}_T(s)$ can be easily computed to determine the error probabilities.

\section{Analysis of Results}
\begin{figure}[h]
\centering{\includegraphics[width=\linewidth]{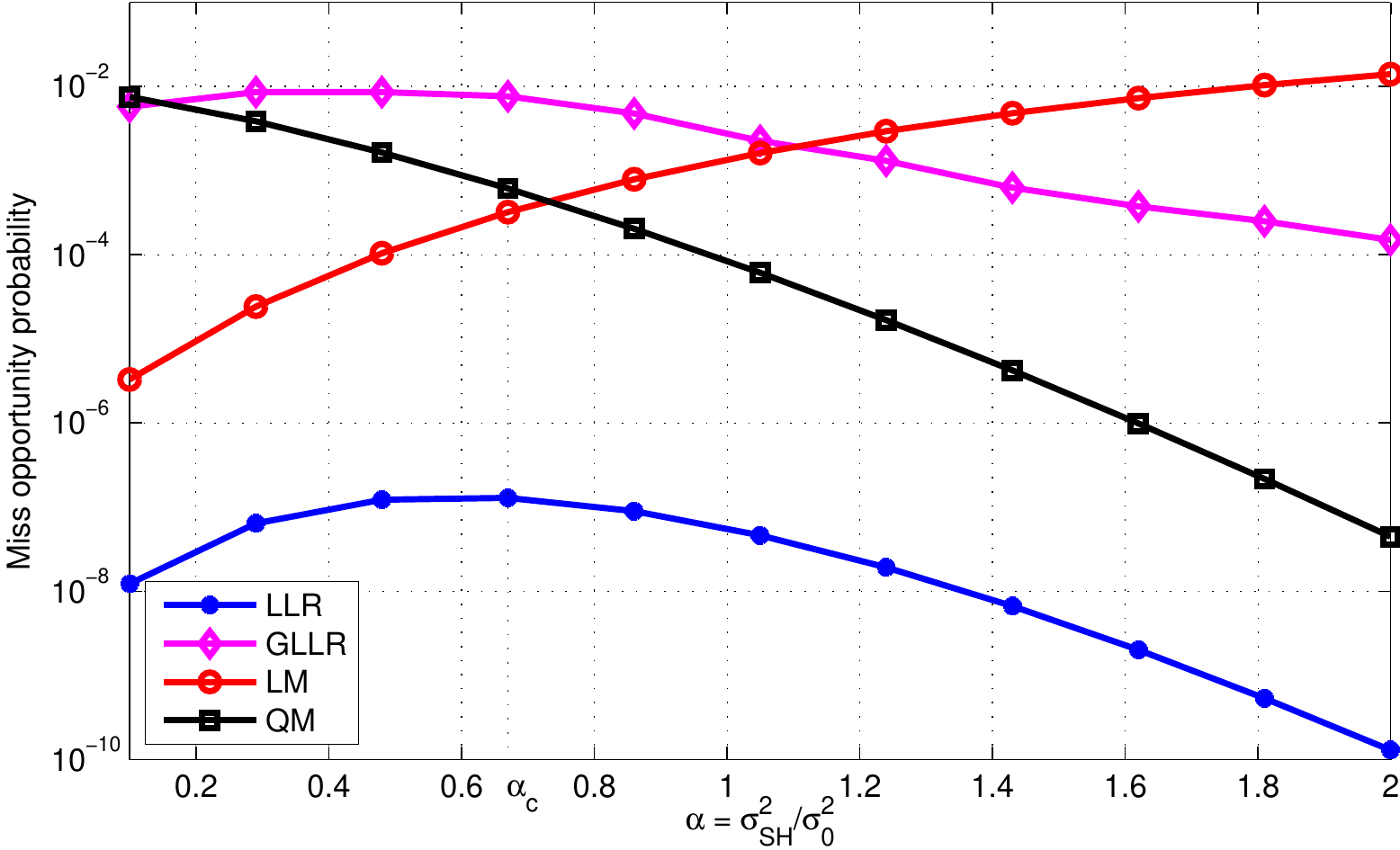}}
\caption{Parameters: $P_\textrm{int}=0.01$, $P_t=.97$ dBm, $d_0=1$, $\gamma=3.3$, $K_\textrm{dB}= 0$ dB, $d_c=0.14$, $n=10$ and $m= 10$.}
\label{fig:PmoValpha}
\end{figure}

In this section, we evaluate the performance of the previous statistics and the optimal LLR statistic, which assumes perfect knowledge of the means and covariance matrices involved in the hypothesis testing problem (\ref{eq:test2}). The LLR provides a lower bound on the miss opportunity probability. 
We consider that the SUs are uniformly distributed in a square of edge $R_s = 0.1$, and that the distance between the center of this square and the PT is $R_p = 1$. We compute the average of the miss opportunity probability $\bar{P}_\textrm{mo}$ subject to $P_\textrm{int}=0.01$ for each realization of the spatial distribution. The error probabilities of the LLR, LM and QM statistics are computed using Thm. 1, while the error probabilities of the GLLR statistic are computed using Monte Carlo simulations. 
The results are shown in Fig. \ref{fig:PmoValpha}, where $\bar{P}_\textrm{mo}$ is plotted against $\alpha=\sigma^2_\textrm{SH}/\sigma^2_0$. Although the LLR performance is several orders of magnitude better that any scheme, it is not feasible in practice given that the true statistics of the measurements are unknown. Let $\alpha_c$ be the cross point of the LM and QM performances. If $\alpha<\alpha_c$, LM performs better, while the converse is true if $\alpha> \alpha_c$. Notice that although the LM and QM statistics were found for the extreme cases when $\alpha\ll 1$ and $\alpha\gg 1$, respectively, both work properly when these conditions are not entirely true.
Notice that if the FC knows in which scenario it is working then, it can select the scheme with best performance that would result in the
minimum of both curves.
Notice also that the GLLR not only requires more energy and bandwidth to transmit their measurements, it also performs worse that the LM or the QM schemes if the FC is aware of $\alpha$ due to estimation errors in $\hat{\ve \mu}$ and $\hat{\tilde{\Sigma}}_1$.
\section{Conclusions}
In this letter, we provided two simple statistics for detecting spectrum wholes in a cognitive radio system. We analyzed them for a practical scenario and observed that their detection performances can be substantially improved with respect to the classical GLLR test if the FC is aware of the shadowing fading variance $\sigma^2_\textrm{SH}$ and the energy detectors inaccuracy $\sigma^2_0$. 

\bibliographystyle{IEEEtran}
\bibliography{IEEEabrv,../Refs/refs}

\end{document}